\documentclass[prd,superscriptaddress,twocolumn]{revtex4}
\usepackage{amstext,amsmath,amssymb,amsfonts,bbm}
\usepackage{amsthm}
\def\be{\begin{equation}}
\def\ee{\end{equation}}
\def\ba{\begin{array}}
\def\ea{\end{array}}
\def\bes{\begin{eqnarray}}
\def\ees{\end{eqnarray}}

\def\6{\langle}
\def\9{\rangle}

\def\1{{{\mathbbm 1}}}

\def\f{\frac}

\def\t {\tilde}

\def\rr{{\cal R}}

\def\hh{{\cal H}}

\def\hP{{\hat{P}}}
\def\hR{{\hat{R}}}
\def\hU{{\hat{U}}}
\def\hV{{\hat{V}}}

\def\hph{{\hat{\phi}}}

\begin{document}

\title{Black hole information problem and quantum gravity}


\author{Daniel R. Terno}
  \affiliation{Centre for Quantum Computer Technology, Faculty of Science, \\ Macquarie University, NSW 2109, Australia}

\begin{abstract}
The gravity-scalar field system in spherical symmetry provides a natural setting for
exploring gravitational collapse and its aftermath in quantum gravity.
In a canonical approach, we give constructions of the constraint and Hamiltonian operators. Matter-gravity entanglement is an inherent feature
of physical states, whether or not there is a black hole. Matter fields alone are an open system with a non-unitary evolution. However, if there is a successful theory of quantum gravity, there is no information loss.
\end{abstract}

\maketitle

%%%%%%%%%%%%%%%%%%%%%%%%%%%%%%%%%%%%%%%%%%%%
%% MAINMATTER
%%%%%%%%%%%%%%%%%%%%%%%%%%%%%%%%%%%%%%%%%%%%

\subsection*{Introduction}

Questions about black hole entropy are a fascinating tangle of general relativity,  quantum field theory, thermodynamics, loops, strings and (quantum) information theory \cite{pt04}.  It was long known that relativity requires some modifications of thermodynamic quantities \cite{ll5}. Both the developments that followed the Bekenstein -- Hawking entropy and holographic principles on the one hand, and a more conventional astrophysical thermodynamics on the other, brought to light various unusual features of the thermodynamics of gravitating systems \cite{buc}.

Black hole evaporation process \cite{bro, fro} makes us to re-examine the basic tenets of quantum mechanics \cite{open}. Application of quantum information theory to the problem of alleged information loss  led to interesting insights and  allowed for a better understanding of different types of entropic quantities. An advantage of using quantum information theory is that it not only provides  a framework to quantify the intuitive ideas of information, but is also applicable when  thermodynamics
 becomes dubious \cite{t04}.

In this contribution I would like to discuss black hole information loss problem in its relationship to a quantum theory.
I begin with  a careful description of  this problem and a summary of its implications. Then I describe a particular approach to tackling it, which is based on a joint work with Viqar Husain \cite{ht}.

\subsection*{Outline and summary}

Following \cite{wald}, the most concise description of the problem is as follows. Black holes evaporate (completely?) via the Hawking process within a finite time. If the correlations between the inside and outside of the black hole are not restored during the evaporation process, then by the time that the black hole has evaporated, an initial pure state will have evolved to a mixed state, i.e., ``information'' will have been lost.

We need to supply a few crucial details. A black hole in question can be treated as arising from the collapse of a matter distribution on the asymptotically  flat spacetime. Hence there is no ``problem of time'' that needs to be resolved. The dynamics, while being constrained, is governed  by a Hamiltonian.

In discussing the apparent paradox the focus is mostly on matter fields on a given (time-dependent) background. The correlations in question   are the ones between the matter field(s) modes, typically inside and outside the horizon. This is consistent with the original derivations of the Hawking radiation and information loss discussion, but misses the fact that matter fields alone are an open system, and as matter -- gravity correlations are established, their entropy can increase. Only a joint treatment of matter and geometry can provide a meaningful answer to this problem \cite{haw05, t05}.

There is an additional aspect that follows from the set-up we just outlined. A classical problem is described by a Hamiltonian, and not by a Hamiltonian constraint. Hence the quantized dynamics has a unitary time evolution, and there cannot be any overall information loss. The increase in the entropy of matter is not an expression of information loss, but a measure of the created entanglement, i.e. redistribution of information. This conclusion is predicated on existence of \textit{some} quantum theory of gravity. \textit{All} current candidates --- loops, strings and foams --- are constructed as unitary theories, and there is no space in them for the information  paradox.

It is also important to realize that for all practical purposes some of the information will be lost. There might be a lower bound on the lost information that is determined by a fundamental physics and not just by experimental or budgetary constraints. Indeed, the fundamental discreteness of space time, as well as quantumness of our measurement devices make us pensive an ideal unitary evolution as non-unitary \cite{gam}. In addition, the reasoning of \cite{mac} is applicable to black holes as well as to a quantum cosmological setting: the state may be pure, but will appear to posses a non-zero entropy.

Part of this entropy results from the unavoidable matter-geometry correlations, which follow from the constraint, as we discuss below. In particular, that forces us to recast the very basic story of the collapse and evaporation: if gravity is taken into account, there is no zero-entropy matter, only a low-entropy one. As a result, the thesis ``pure states do not wear black'' gets an additional meaning.

\subsection*{Classical setting}

  Our starting point   is the Arnowitt-Deser-Misner (ADM) hamiltonian formulation for general relativity.
 The phase space of the model is defined by prescribing a
form of the gravitational phase space variables $q_{ab}$ and
$\tilde{\pi}^{ab}$, together with fall-off conditions  for these
variables, and for the lapse and shift functions $N$ and $N^a$.  The
bulk  ADM 3+1 action for general relativity minimally coupled to a
massless scalar field is
\be
S = \frac{1}{16\pi G}\int d^3x dt\left[ \tilde{\pi}^{ab}\dot{q}_{ab} +
\t{P}_\phi\dot{\phi}
- N H - N^a C_a\right].
\label{act}
\ee
The pair $(\phi,P_\phi)$ are the scalar field canonical variables, and the hamiltonian and spatial diffeomorphism constraints take their standard form.
This action (together with the boundary terms, see e.g. \cite{hw-flat, eric}) is well-defined and determines the fall-off conditions on canonical variables.
  The reduction to spherical symmetry utilizes an  auxiliary
flat Euclidean metric $e_{ab}$ and unit radial normal  $s^a= x^a/r$, where  $r^2=e_{ab}x^ax^b$.
The parametrization is given by two geometric dynamical variables, $\Lambda(r,t)$ and $R(r,t)$, and their canonical conjugates. Hence a spatial line element is given by
\be
ds^2 = \Lambda^2(r,t)  dr^2 + R(r,t)^2 d\Omega^2.
\ee

The Painleve-Gullstrand  (PG) coordinates are those  where equal coordinate time slices are
spatially flat \cite{hw-flat,eric}.  It is sufficient to use the partial gauge fixing
$\Lambda=1$  to obtain  non-singular coordinates at the horizon, which is
the feature of PG coordinates we desire.

The ADM $3+1$ action with minimally coupled
scalar field  leads to the reduced action and the reduced Hamiltonian and  radial diffeomorphism constraints.
These   constraints are first class with an algebra that is similar to that of the full theory  \cite{hw-flat}.

We next impose the  gauge choice $\Lambda=1$, which corresponds to a step toward flat slice coordinates.   With this gauge condition, the Hamiltonian constraint is solved (strongly) for the conjugate momentum $P_\Lambda$ as a function of the phase space variables.  This gives
\be
P_\Lambda = P_RR + \sqrt{ (P_RR)^2 - X},
\label{PL}
\ee
where
\be
X =  16R^2 (2RR'' - 1 + R'^2) + 16R^2 H_\phi,
\ee
and
\be
 H_\phi = \frac{P_\phi^2}{2R^2} + \frac{R^2}{2}\ \phi'^2.
\ee
The evolution equation for $\Lambda$ \cite{hw-flat} and the requirement that the gauge $\Lambda=1$ be preserved under it leads to fixing of the
lapse $N$ as a function of the shift $N^r$.
Finally the reduced gravitational Hamiltonian is
\begin{align}
H_{R}^G& =\int_0^\infty (N^r)'\left( R P_R + \sqrt{(P_RR)^2 - X}\right) dr \nonumber \\
&+ \int_0^\infty N^r(P_RR' + P_\phi \phi')\ dr,
\label{Hred}
\end{align}
where the surface term in the reduced action has been written as bulk term and combined
with the remaining radial diffeomorphism constraint.  This is the reduced system we quantize.

\subsection*{Quantization}
We use the polymer quantization \cite{hw_quant} which may be viewed as the ``dual''   to that used in loop quantum gravity. Its  basic variables are
\be
R_f = \int_0^\infty dr  f(r) R(r), \qquad  U_\lambda(P_R)= \exp\left(i\lambda P_R\right),
\label{basic-var}
\ee
where $f(r)$ is a smearing function and $\lambda$ is a  real constant. We use  similar
definitions for the variables made from $\phi$ and $P_\phi$. These satisfy the canonical Poisson bracket
\be
\{ R_f , U_\lambda(P_R(r)) \}  = i \lambda f(r) U_\lambda(P_R(r)).
\label{basicpb}
\ee

This Poisson bracket is  realized as an operator relation on a Hilbert space with basis states
$| a_1,a_2,\cdots ,a_n\rangle$,
where the real numbers $a_i$ represent the radial field ($R$) values at the radial lattice points $r_i$. The inner
product  is
\be
\langle a_1',a_2',\cdots ,a_n'| a_1,a_2,\cdots ,a_n\rangle= \delta_{a_1',a_1}\cdots \delta_{a_n',a_n},
\ee
if two states are associated with the same  lattice points; if not the inner product is zero.  For our purpose it suffices to consider a fixed  lattice, so the latter situation does not arise.  This inner product is background independent in the same way  as for example
the inner product for the Ising model; the difference is that for the latter there is a finite dimensional spin
space at each lattice point.
With the inclusion of matter the kinematical Hilbert space is the  tensor product of geometry
and matter Hilbert spaces with basis
\be
|\underbrace{a_1,\ldots, a_N}_{gravity};\underbrace{b_1,\ldots,
b_N}_{matter}\9,
\ee
In this quantization there are three lattices:   the discretized three-manifold (the radial
points $r_1\cdots r_k$) with spacing $ l_P\lambda$, where $l_P$ is the Planck length, the gravity configuration lattice with spacing $\epsilon$, and the matter configuration lattice $\kappa$.

A localized field  may be defined by taking for example $f(r)$ in Eq. (\ref{basic-var}) to be a Gaussian
or a  smooth function of a bounded support which sharply peaked ($\sigma \ll 1$ in the Gaussian case)  at a radial point $r_k$. We   assume this in the following and set $R_{G(r_k)}\equiv R_k$,
 etc. With this convention the basic geometry operators act as
\begin{align}
& \hR_k|{a_1,\ldots, a_N};{b_1,\ldots,
b_N}\9 \nonumber \\
& =2l_P^2a_k|{a_1,\ldots, a_N};{b_1,\ldots, b_N}\9,
\end{align}
\begin{align}
& \hU_k(\epsilon)|{a_1,\ldots, a_N};{b_1,\ldots,b_N}\9 \nonumber \\ &=|{a_1,\ldots,a_k-\epsilon,\ldots a_N};{b_1,\ldots,
b_N}\9
\end{align}
and the field operators act similarly on the corresponding degrees of freedom.

It is readily verified that the commutator of these operators is a faithful realization of the corresponding Poisson bracket.  The parameter $\lambda$ represents the discreteness scale in field configuration
space.  There are similar definitions of the basic operators for the matter sector.

This representation is one in which the momentum operator does not exist. There is however
an alternative $\epsilon$ dependent definition of momentum using the translation operators,
 given by
\be
\hat{P}_R^\lambda(r_k) := \frac{1}{2i\epsilon}\ \left(\hat{U}_\epsilon(P_R(r_k)) - \hat{U}_\epsilon^\dagger P_R((r_k)) \right).
\label{mom}
\ee

With the above choice of the smearing functions the commutators are
\be
[\hR_k,\hU_l(\epsilon)]=-2l_P^2\epsilon\delta_{kl}\hU_l(\epsilon),
\ee
and
\be
 [\hph_k,\hV_l(\kappa)]=-2l_P^2\kappa\delta_{kl}\hV_l(\kappa)
\ee

Since $\hR_k$ operators have zero in the spectrum (ie. there are states such as $|a_1 \cdots, a_k=0,\cdots a_n\rangle$
that have zero eigenvalue), there is no  inverse operator and an indirect definition is required. This is achieved by
Poisson bracket identities such as
\be
\{\sqrt{|R_k|},U_l(\epsilon)\}=\f{1}{2\sqrt{|R_k|}}i\epsilon
U_l(\epsilon)\delta_{kl},
\ee
first noted by Thiemann.  The operator
\be
\widehat{\f{1}{R_k}}\equiv\left(\f{2}{2l_P^2\epsilon^2}\hU_k(-\epsilon)\left[\sqrt{|\hR_k|},\hU_k(\epsilon)\right]\right)^2
\ee
depends on the field lattice spacing $\epsilon$ through its dependence on the translation operators $\hU_k$.
 It it is diagonal in the basis $|a;b\9$.

 Turning to the terms with spatial derivatives we can define operators using the  convection
 \be
f'(r)\rightarrow \f{f_{k+1}-f_{k}}{l_P\lambda}, \qquad f''(r)\rightarrow\f{ f_{k+1}-2f_k+f_{k-1}}{2l_P^2\lambda^2},
\ee
for any function $f_k$. Lastly the lattice local observable momentum operators  is
\be
\hP_{Rk}\equiv\f{l_P}{2i\epsilon}(\hU_k(\epsilon)-\hU_k^\dag(\epsilon)),
\ee
the squared momentum operator
\be
\hP_{Rk}^2\equiv\f{l_P^2}{\epsilon^2}(2-\hU_k(\epsilon)-\hU_k^\dag(\epsilon)),
\ee
and similar expression for the matter field momentum.

In constructing more complicated operators the question of the operator ordering is important. One option is a symmetric ordering.
Another possibility is an order at which, e.g., $\hR$ is to the right of $\hP_R$. It has an advantage of annihilating the state of zero gravitational excitations.

Our next goal is to use the above definitions of basic operators to give a prescription for the
radial diffeomorphism constraint and the reduced Hamiltonian (\ref{Hred}).  The main issue is the definition of the square root $\sqrt{Y}$ in this Hamiltonian,
where
\be
Y=(P_RR)^2 -\rr- 16R^2 H_\phi, \qquad \rr=16R^2(2RR'' - 1+ R'^2).
\ee
Substituting for $P_\Lambda$ and computing its derivative we find
\be
C_r=-P'_RR-\frac{Y'}{2\sqrt{Y}}+P_\phi\phi'\simeq 0
\ee
If supplemented with the requirement $Y>0$ (which is the case for classical solutions), this is equivalent to
\be
C=Y'^2/4-(P_\phi\phi'-P'_RR)^2Y\simeq 0
\label{norootdiff}
\ee
In this form it does not contain a square root and it is now straightforward to construct the corresponding
operator using the basic ones defined above.

The Hamiltonian is a surface term for which we need to define an operator for $P_\Lambda$.  A direct quantization of the Hamiltonian \be
\hat{H}^{\mathrm{phys}}_k={N^r}_k'(\widehat{P_RR})_k+|\hat{Y}_k|^{1/2}+N^r_k\left((\widehat{P_R'R})_k+(\widehat{P_\phi
\phi'})_k\right),
\ee
requires to define the action of $|\hat{Y}_k|^{1/2}$. Since $\hat{Y}_k$ is not diagonal
in the basis we are using, this operator is not easy to define unless we go to a different basis.  However,
from the constraint $C_r$ we see that classically on the constraint surface we have
\be
P_\Lambda(r)  = \int ^r_0 dr' (P_\phi \phi' + P_RR').
\ee
This suggests that for physical states it is possible to compute the energy by finding an operator analog of the
r.h.s. of this equation. Since the quantization we are using utilizes a radial lattice we can write the integral as a discrete
sum over the lattice points  $r_k$. It is therefore reasonable to suggest the definition
\be
\hat{P}_{\Lambda k} =  \lambda l_P\sum_{i=1}^k  \left[ (\widehat{P_\phi \phi'})_i + (\widehat{P_R R'}_i )\right]
\ee
In order to use this operator we would first need to find physical states $|\psi\rangle_{P}$ , ie. states that are annihilated by the operator analog  of the constraint (\ref{norootdiff}), and then compute $\ _P\langle \psi | \hat{P}_{\Lambda k}| \psi \rangle_P$. The energy of the quantum spacetime would be this expression evaluated at the farthest lattice point from the origin, in keeping with the classical
definition where the energy
\be
E= \lim_{r\to \infty}\ P_\Lambda(r)N^r(r).
\ee
 We note that the asymptotic falloff of $N^r$ is determined by the
classical requirement of functional differentiability \cite{hw-flat}, and this behavior of $N^r$ carries over to the quantum theory.

\subsection*{Physical states and entanglement}

Initial states of the gravity-matter system should satisfy the quantum constraint
\be
\hat{C}|\psi\9=0,
\ee
which is supplement by the requirement that $|\psi\9$ belongs to the positive part of the spectrum of $|\hat{Y}|$. It remains to be seen wether any of the operator ordering choices allows for this to be satisfied on a sufficiently large set of states, or if another  realization of the constraint is necessary to accomplish this.  Nevertheless, it is already possible to make a few remarks.

Firstly,   in the case of pure gravity $(\phi=P_\phi=0)$, the ordering that puts the $\hR$ operator to the right results in  the operator form  of the constraint  (\ref{norootdiff}) gives
\be
\hat{C}|0\9=\left(\hat{Y}'^2/4-\hat{P}'^2_R\hat{R}^2\hat{Y}\right)|0\9=0,
\ee
where $|0\9$ stands for the state with no excitations. This may be viewed as a ``degenerate metric vacuum'' because
it is the eigenstate of the field operator $\hat{R}_k$   with zero eigenvalue at all points $r_k$.

Secondly, the presence of the matter-gravity terms in the constraint turns it into entangling operator \cite{pt04}.
 This can be seen by considering the constraint  the form of Eq.~(\ref{norootdiff}). It splits into geometry ``interaction'' matter--geometry parts, since  the way we defined $C$ removes a matter term,
\be
\hat{C}=\hat{C}_G\otimes\1_M+\sum_\alpha\hat{C}^G_\alpha\otimes\hat{C}^M_\alpha. \label{constrsplit}
\ee
The latter term contains monomials that involve gravity and matter operators. Consider their action on the basis states of  the kinematical Hilbert space $\hh=\hh_G\otimes\hh_M$. The term $\hat{P}_{\phi k} \hat{P}_{R k}$ serves as an example. Its action results in
 a direct product of entangled gravity and matter modes. However, adding different terms in the constraint results in a superposition of such direct product states, and the constraint operator cannot be written as $\hat{C}=\hat{C}_G\otimes\hat{C}_M$. Some states of the form $|\psi\9_G\otimes|\varphi\9_M$ may remain in  a direct product form,
% \vspace{-1mm}
  \be
 \hat{C}\left(|\psi\9_G|\varphi\9_M\right)=|\psi'\9_G|\varphi'\9_M.
 \ee
 However, they do not span a linear space (of states that remain direct product under the action of $\hat{C}$), since their superpositions are by definition entangled. Moreover, imposition of the constraint reduces this set even further, since now
 \be
 \sum_\alpha\hat{C}^G_\alpha\otimes\hat{C}^M_\alpha\left(|\psi\9_G|\varphi\9_M\right)\propto |\psi'\9_G|\varphi\9_M,
 \ee
must hold as well.

This  entanglement of the gravitational and matter excitations  prevents the direct product gravity-matter decomposition of a generic physical state. It remains to be seen weather  any physical states other than the ``vacuum''  $|0\rangle$ exist at all. As a result, the most basic formulation of the information loss paradox ``a pure state goes to a mixed state'' is physically untenable because it ignores the entanglement generically present already at the initial state stage.
The same conclusion ---  ``pure states do not wear black'' was argued for also in string theory \cite{myers}.

\bibliography{}

\end{document}